\documentclass{article}

\usepackage{wrapfig}
\usepackage[portuges,english]{babel}
\usepackage{amssymb,latexsym,amsmath,color,mathrsfs,ifsym,graphics,stmaryrd}
\usepackage{amsthm, amsfonts, mathrsfs}
\usepackage[colorlinks,linkcolor=blue,urlcolor=blue,citecolor=black,
plainpages=false,pdfpagelabels,breaklinks]{hyperref}
\usepackage[all]{xy}

\usepackage{graphicx}
\usepackage{caption}
\usepackage[margin=2 cm]{geometry}

\DeclareMathOperator{\EA}{EA}

\begin{document}

\title{{\sc Everything is Entangled in Quantum Mechanics:\\Are the Orthodox Measures Physically Meaningful?}}

\author{{\sc Christian de Ronde}$^{1,2,3}$, {\sc Raimundo Fern\'andez Mouj\'an}$^{2,4}$, {\sc C\'esar Massri}$^{5,6}$}
\date{}

\maketitle
\begin{center}
\begin{small}
1. Philosophy Institute Dr. A. Korn, University of Buenos Aires - CONICET\\
2. Center Leo Apostel for Interdisciplinary Studies\\Foundations of the Exact Sciences - Vrije Universiteit Brussel\\
3. Institute of Engineering - National University Arturo Jauretche\\
4. Philosophy Institute, Diego Portales University, Santiago de Chile\\
5. Institute of Mathematical Investigations Luis A. Santal\'o, University of Buenos Aires - CONICET\\
6. CAECE University
\end{small}
\end{center}

\begin{abstract}
\noindent Even though quantum entanglement is today's most essential concept within the new technological era of quantum information processing, we do not only lack a consistent definition of this kernel notion, we are also far from understanding its physical meaning \cite{Horodecki09}. These failures have lead to many problems when attempting to provide a consistent measure or quantification of entanglement. In fact, the two main lines of contemporary research within the orthodox literature have created mazes where inconsistencies and problems are found everywhere. While the operational-instrumentalist approach has failed to explain how inequalities are able to distinguish the classical from the quantum, the geometrical approach has failed to provide a consistent meaningful account of their entropic measure. Taking distance from orthodoxy, in this work we address the quantification and measure of quantum entanglement by considering a recently presented objective-invariant definition in terms of the coding of intensive relations \cite{deRondeMassri21b} which allows to escape the widespread relativist account of bases and factorizations \cite{deRondeMassri23, deRondeFMMassri23a}. Going beyond the orthodox dualistic reference to ``quantum particles'' and ``clicks'' in detectors, we will argue that this new line of research is capable not only to evade the many open problems which appear within the mainstream literature, but is also able to present a consistent and coherent physical understanding of entanglement. The main conclusion of this work is that in quantum mechanics --contrary to what is generally presupposed-- all operational expressions found within the laboratory are intrinsically entangled. 
\end{abstract}
\begin{small}

\end{small}

\newtheorem{theorem}{Theorem}[section]
\newtheorem{definition}[theorem]{Definition}
\newtheorem{lemma}[theorem]{Lemma}
\newtheorem{proposition}[theorem]{Proposition}
\newtheorem{corollary}[theorem]{Corollary}
\newtheorem{remark}[theorem]{Remark}{\hspace*{4mm}}
\newtheorem{example}[theorem]{Example}

\bigskip

\bigskip

\bigskip

\bigskip

\section*{Introduction}


The notion of quantum entanglement was born as a critique, as a sort of \emph{reductio ad absurdum}, of the Standard formulation of Quantum Mechanics (SQM), whose main architect was undoubtedly Niels Bohr. First Einstein, in the famous EPR paper \cite{EPR}, and then Schrödinger in his also famous ``cat paper''  \cite{Schr35a} seemed to show some absurd consequences when considering physical reality in terms of SQM. If the famous ``collapse'' of quantum superpositions introduced by Dirac was to be taken seriously, there appeared to be an impossible action at a distance between ``particles'' which defied the limit imposed by the velocity of light. But Bohr's immediate reply --even though no one really understood it-- would be taken by the physics community as the final triumph of the Danish physicist --and his complementarity approach-- over an old senile Einstein, who had failed to understand the revolution that was taking place. The young notion of entanglement was then buried alive and soon forgotten by the new instrumentalist orthodoxy. However, three decades later, going back to the forbidden works of Einstein and Schrödinger, John Bell would rediscover, in secrecy, a set of testable inequalities capable of attacking the problem. His investigations, published in a marginal Journal, would remain completely unnoticed by the physics community until, one decade later, a young student called John Clauser would encounter by pure luck Bell's papers. Regardless of the instrumentalist context that had banned these foundational investigations, Clauser would commit himself to finding out if QM did or did not violate the inequalities that Bell had published. Performing these experimental test in 1978 would, of course, cause the young Clauser a lot of difficulties for finding a permanent position within the academic world (see \cite[pp. 254--255]{Becker18}), but this persecution would not stop the revolts against orthodoxy. Alain Aspect --who did posses a permanent position in the academic world-- would continue the tests, together with his group in France, Bell inequalities confirming the violation. During the 1980s, all these works would become an essential fuel for a new field of research called ``Philosophy of Quantum Mechanics'' where physicists, mathematicians, logicians and philosophers would gather to openly discuss the many problems of SQM. Finally, during the 1990s, leaving aside its critical origin,  instrumentalist physicists would come to accept the utility of entanglement reframing the notion in terms of the orthodox narrative. Turning the problem into a solution, without further analysis, physicists were ready to subvert the notion of entanglement and present it --although without a real understanding-- as part of their own algorithmic recipes. As we will see, this would lead to an inconsistent account of one of the most essential notions of contemporary physics. 

In the present work we will focus on the measure and quantification of quantum entanglement. In order to do so we will provide a critical analysis of the way in which the orthodox contemporary literature has addressed the definition of entanglement in essentially inconsistent terms. We will then discuss the two main lines of research which have attempted to define different measures of entanglement. After these critical considerations we will present a completely different account to entanglement grounded on the logos categorical approach to QM. Going beyond the orthodox reference to ``quantum particles'' and ``clicks'' in detectors, we will argue that this new line of research is capable not only to evade the many open problems which appear within the orthodox literature, but also to present a consistent and coherent physical understanding of the quantification of entanglement. The main conclusion of this work is that in quantum mechanics --contrary to what is generally presupposed-- each and every state of affairs described by the theory is intrinsically entangled. Or in other words, that there is nothing non-entangled within the theory of quanta.

\section{The Contemporary (Inconsistent) Account of Entanglement}

Today, entanglement is unanimously accepted and it is commonly argued that it represents an ``holistic property of compound quantum systems, which involves nonclassical correlations between subsystems'' \cite[p. 865]{Horodecki09}. In the introduction to the book, \emph{Philosophy of Quantum Information and Entanglement}, the concept is presented in the following manner:
\begin{quotation}
\noindent {\small ``Consider two particles, $A$ and $B$, whose (pure) states can be represented by the state vectors $\psi_A$ and $\psi_B$. Instead of representing the state of each particle individually, one can represent the composite two-particle system by another wavefunction, $\Psi_{AB}$. If the two particles are unentangled, then the composite state is just the tensor product of the states of the components: $\Psi_{AB} = \psi_A \otimes \psi_B$; the state is then said to be factorable (or separable). If the particles are entangled, however, then the state of the composite system cannot be written as such a product of a definite state for $A$ and a definite state for $B$. This is how an entangled state is defined for pure states: a state is entangled if and only if it cannot be factored: $\Psi_{AB} \neq \psi_A \otimes \psi_B$.'' \cite[p. xiii]{BokulichJaeger10}} 
\end{quotation}   
However, if we analyze this definition from a critical point of view, following Einstein and Schrödinger’s attitude, we come to see that the notions of particle, separability and purity, commonly applied within SQM, are essentially inconsistent, precluding the possible understanding of entanglement right from the start. These facts have remained almost unnoticed to the general physics community, which tends to regard its own discourse as devoid of any consistent reference, claiming --specially when problems pop up-- that all this is ``just a way of talking'' (see for a detailed discussion \cite{deRondeFM21, deRondeMassri20b}).

\subsection{The Myth of Quantum Particles}

One of the main cornerstones of Bohr’s program is the combination of, on the one hand, a narrative according to which QM talks about a ``microscopic'' physical realm constituted by ``elementary particles'', and, on the other hand, the idea that these elementary particles escape theoretical representation, that it is impossible to conceptually apprehend this postulated microscopic realm. In this manner, Bohr astutely accomplished the imposition of an atomist narrative while at the same time he convinced physicists of the irrepresentability of quantum particles themselves, precluding in this way the possibility of any critical revision. The way in which made up fictions such as ``quantum particles'' and ``quantum jumps'' have been justified --quite regardless of any theoretical representation or experimental evidence-- in the context of SQM was exposed by Werner Heisenberg in his autobiography. The German physicist was a direct witness of a meeting between Bohr and Erwin Schr\"odinger, which took place in Copenhagen in 1926, in order to discuss the existence (or not) of ``quantum jumps''. As Heisenberg would recall, even though the many arguments that Schr\"odinger \cite[p. 73]{Heis71} had produced during the debate had allowed him to rationally conclude that ``the whole idea of quantum jumps is sheer fantasy'' the Danish illusionist, with a single move of his magic wand --a term that Arnold Sommerfeld had used to characterize Bohr's introduction of the correspondence principle--, would invert the burden of proof turning things completely upside-down: 
\begin{quotation}
\noindent {\small ``What you say is absolutely correct. But it does not prove that there are no quantum jumps. It only proves that we cannot imagine them, that the representational concepts with which we describe events in daily life and experiments in classical physics are inadequate when it comes to describing quantum jumps. Nor should we be surprised to find it so, seeing that the processes involved are not the objects of direct experience.'' \cite[p. 74]{Heis71}} 
\end{quotation}
This is a great example of Bohr’s tactics, as just explained: he postulated the existence of ``quantum jumps'', and then rejected Schrödinger’s critique of these postulated phenomena by referring to their irrepresentability. Confessing his impotency, Schrödinger would write to his friend Wilhelm Wien: 
\begin{quotation}
\noindent {\small ``Bohr's [...] approach to atomic problems [...] is really remarkable. He is completely convinced that any understanding in the usual sense of the word is impossible. Therefore the conversation is almost immediately driven into philosophical questions, and soon you no longer know whether you really take the position he is attacking, or whether you really must attack the position he is defending.'' \cite[p. 228]{Moore89}} 
\end{quotation}   
In any case, this has helped to uncritically retain the fundamentally unjustified claim that QM talks about elementary particles such as electrons, protons and neutrons. This atomist presupposition is still today one of the main obstacles for the understanding of the theory of quanta. One may take that idea as an exaggeration, claiming that quantum particles are ``just a way of talking''. But, in fact, the problem is deeper, as the atomist supposition implies, wether we like it or not, a series of deductions, and methodological and operational steps, that determine right from the start the understanding of the formalism and of observations. As Faraday explained long ago: ``the word \emph{atom}, which can never be used without involving much that is purely hypothetical, is often intended to be used to express a simple fact; but good as the intention is, I have not yet found a mind that did habitually separate it from its accompanying temptations'' \cite[p. 220]{Laudan81}. Schrödinger rephrases this idea for the quantum case: ``We have taken over from previous theory the idea of a particle and all the technical language concerning it. This idea is inadequate. It constantly drives our mind to ask information which has obviously no significance'' \cite[p. 188]{Schr50}. It is not difficult to understand that if one dogmatically applies a series of categorical principles --such as those of particle metaphysics, e.g., separability, individuality, locality, etc.-- to a mathematical formalism that was never meant to be understood under the constraints of such representation, the result of this methodology will lead only to paradoxes and dead ends. In fact, as we will later show, the birth of the quantum formalism through matrix mechanics was due to Heisenberg's decision to abandon the atomist discourse, to forget about the trajectory of presupposed particles, and to concentrate on the intensive quantities that were, in fact, actually observed in the lab. It was this attention to intensities, gained by the abandonment of the atomist narrative, that allowed to construct an invariant mathematical formalism. But, since physicists were so attached to the atomist worldview common to classical physics, they were not ready to accept the idea of a physical element of an intensive nature, and they projected once again a world of particles into the quantum formalism, destroying the invariance that was present in Heisenberg's matrix mechanics and, for the first time in the history of physics, embracing perspectival relativism as acceptable (we will come back to this essential point in section 3). 

During the 1990s, after entanglement was finally awakened from its half century hibernation, it is this same atomist narrative which was applied by instrumentalist trained physicists in order to ``explain'' the concept. As a consequence, today, most papers about quantum entanglement begin with the explicit reference to “quantum particles”. Only in some cases, an attempt to avoid the reference to particles is made through the euphemistic reference to quantum \emph{systems} and \emph{subsystems}. But a simple change of words, without a comprehensive critical analysis, is not enough, as those deductive, methodological and operational steps, originated by the atomist discourse, continue to function. In any case, the reference to particles is present --explicitly or implicitly-- in the introduction to almost every published paper about quantum entanglement. Just to give a few examples coming from some of the most prestigious researchers in the field:
\begin{itemize}
\item  Davide Castelvecchi and Elizabeth Gibney \cite{Nature22}: ``Because of the effects of quantum entanglement, measuring the property of one particle in an entangled pair immediately affects the results of measurements on the other. It is what enables quantum computers to function: these machines, which seek to harness quantum particles' ability to exist in more than one state at once, carry out calculations that would be impossible on a conventional computer.'' 
\item Richard Cleve and Harry Buhrman \cite{CleveBuhrman97}: ``If a set of entangled particles are individually measured, the resulting outcomes can exhibit `nonlocal' effects. These are effects that, from the perspective of `classical' physics, cannot occur unless `instantaneous communications' occur among the particles, which convey information about each particle's measurement to the other particles.''
\item Ryszard, Pawe, Micha and Karol Horodecki \cite{Horodecki09}: ``[Entanglement is an] holistic property of compound quantum systems, which involves nonclassical correlations between subsystems.''
\item Jian-Wei Pan, Dik Bouwmeester, Harald Weinfurter and Anton Zeilinger \cite{Zeilinger98}: ``entanglement has been realized either by having the two entangled particles emerge from a common source, or by having two particles interact with each other. Yet, an alternative possibility to obtain entanglement is to make use of a projection of the state of two particles onto an entangled state.''
\item Abner Shimony \cite{Shimony95}: ``A quantum state of a many-particle system may be `entangled' in the sense of not being a product of single-particle states.'' 
\item Thomas, R.A., Parniak, M., Ostfeldt, C. et al. \cite{Thomas21}: ``Entanglement is an essential property of multipartite quantum systems, characterized by the inseparability of quantum states of objects regardless of their spatial separation.''
\item Vlatko Vedral \cite{Vedral14}: ``entanglement can exist in many-body systems (with arbitrarily large numbers of particles).'' 
\item William K. Wootters \cite{Wootters98}: ``Quantum mechanical objects can exhibit correlations with one another that are fundamentally at odds with the paradigm of classical physics; one says that the objects are `entangled'.''
\end{itemize}

But this paradoxical construction does not stop here. There are other inconsistencies in the orthodox notion of entanglement that can be also found when analyzing the notions of `purity' and `separability'  (essentially related to the atomist supposition), to which we will now turn our attention.

\subsection{The Inconsistency of Purity}\label{section-pure}

As it has been exposed in \cite{deRondeMassri22a}, the notion of purity introduced in SQM is an essentially inconsistent notion grounded on the idea that QM is about predicting single `clicks' in detectors --which are supposedly consequence of microscopic particles-- in a certain binary manner. This binary understanding, based on the atomist presupposition, has destroyed the invariance found --as we will show-- in Heisenberg's matrix mechanics, and has precluded the possibility of a consistent global valuation of a given state of affairs in QM. And, furthermore, since Dirac's account of quantum states, there has been, within the orthodox literature,  a self-contradictory double-reference to `pure states' which has not been recognized. While on the one hand there exists a non-invariant operational definition in terms of the possibility to predict with certainty (with probability = 1) single measurement outcomes (taken as consequence of small corpuscles), there is also a reference to pure states linked to abstract vectors which, in fact, have no operational content whatsoever. As it has been demonstrated in \cite{deRondeMassri22a} these two definitions are not equivalent nor consistent. This leads to a widespread confusion where two \emph{different} states, considered from the standpoint of the operational definition in terms of certain outcomes, can be, from the point of view of the abstract definition, taken as the \emph{same} state.

Dirac and von Neumann were both mathematicians, not physicists, and this might have been the reason behind their misuse of the notion of {\it state} in terms of {\it abstract vectors} on the one hand, and in terms of {\it basis represented vectors} (i.e., {\it kets}) on the other. The deep and problematic redefinition of the meaning of state in QM was recoginzed by Pauli and Schr\"odinger. Arthur Fine \cite[p. 94]{Schlosshauer11} comments: ``Wolfgang Pauli thought that using the word `state' ({\it Zustand}) in QM was not a good idea, since it conveyed misleading expectations from classical dynamics.'' Erwin Schr\"odinger \cite[p. 153]{Schr35a} would also criticize Dirac's use of the notion of state arguing that: ``The classical concept of {\it state} becomes lost [in QM], in that at most a well-chosen {\it half} of a complete set of variables can be assigned definite numerical values.'' As discussed in detail in \cite{deRondeMassri22a}, the essential inconsistency present in Dirac's re-definition of (quantum) state is related to the misuse of {\it reference frames} as a precondition to account consistently for any operational reference within {\it the same} physical situation. Operationality is undoubtedly a necessary characteristic of physical concepts. Something that was remarked by Einstein: 
\begin{quotation} \noindent {\small ``The concept does not exist for the physicist until he has the possibility of discovering whether or not it is fulfilled in an actual case. We thus require a definition of [the concept] such that this definition supplies us with the method by means of which, in the present case, he can decide by experiment whether or not [the concept] occurred.'' \cite[p. 26]{Einstein20}} \end{quotation}
But the operational content of a theory must be invariant when attempting to address the same state of affairs throughout different reference frames. That is of course the whole point of invariance. It is operational-invariance which allows to represent consistently the same state of affairs independently of any particular reference frame. However, it is in order to save a binary observational reference to single measurement outcomes (consequence, supposedly, of particles), that we find, in the operational definition presented by Dirac, the sacrifice of invariance. This is explicit in the operational definition of the notion of pure state commonly applied in the orthodox literature: if a quantum system is prepared in a {\it maximal basis} so that there is a {\it maximal test} yielding with certainty a particular outcome, then it is said that the quantum system is in a \emph{pure state}. As it is well known since Kochen and Specker's explicit demonstration \cite{KS}, the operational content within this preferred reference frame (or basis) cannot be invariantly translated into another reference frame (regarding the same state of affairs). There is only one basis in which such binary certainty can be actually obtained for an abstract vector, something which in Dirac's notion is given in terms of any one term {\it ket}, $| x \rangle$. Thus, the invariance of the formalism (as well as the consistency of the notion of state) is lost, replaced by a perspectival relativism that implies the choice of a particular basis.
\begin{definition}[Operational Purity] Given a quantum system in the state $|\psi \rangle$, there exists an experimental situation linked to that basis (in which the vector is written as a single term) in which the test of it will yield with certainty (probability = 1) its related outcome.\footnote{Von Neumann's application of this notion in the context of quantum logic is also explicit as related to his definition of {\it actual property}, something applied in the many operational approaches that were developed during the 1960s and 1970s (see for a detailed analysis \cite{deRondeFreytesDomenech18}). In short, a property is {\it actual} if given a specific experimental set up we know with certainty (probability = 1) the result of the future outcome (see also \cite{Aerts81, Piron76}).} 
\end{definition}
\noindent  
This definition precludes the consistent translation (of the operational content) between different basis dependent accounts of a state. Consequently, the possibility to refer to {\it the same} state independently of reference frames is lost. But there is also a co-existent widespread definition of pure states in basis-independent terms: any {\it abstract vector} is defined as an invariant element under rotations. Thus, any rotation of a pure state, now understood as a vector in purely abstract terms (i.e., independent of any basis) must be considered to be {\it the same} state.  
\begin{definition}[Abstract Purity]\label{pure} An abstract unit vector (with no reference to any basis) in Hilbert space, $\Psi$, is a pure state. In terms of density operators $\rho$ is a pure state if it is a projector, namely, if Tr$(\rho^2) = 1$ or $\rho = \rho^2$. 
\end{definition}
\noindent It is at this point that we need to clearly distinguish between a purely abstract vector, $\psi$, and its specific representations in different bases, as, for example, $|\psi\rangle$ or $c_1 |\phi_1\rangle + c_2 |\phi_2\rangle$.\footnote{As shown in detail in \cite[Sect. 4]{deRondeMassri19a} this distinction becomes explicitly visualizable through the use of graph theory (see figures 1 and 6 of the mentioned reference).} As discussed in detail \cite{deRondeMassri22b}, this notation helps to understand the essential equivocity present within the literature where these two (inconsistent) definitions of purity are used and applied interchangeably. But while {\it operational purity} is basis-dependent (i.e., it explicitly depends on a {\it maximal basis} were the abstract vector $\psi$ is written as a single term ket $|\psi\rangle$) and consequently non-invariant, {\it vectorial purity} (i.e., the reference to the abstract invariance of $\psi$) provides an invariant definition but has no operational content whatsoever. Thus, while the operational definition destroys the invariant reference of the mathematical formalism, the latter abstract definition lacks an experimental counterpart. Furthermore, there is no equivalence between these two distinct definitions. While operational purity implies vectorial purity the converse is false. An abstract vector does not imply a specific basis-dependent representation. As we will discuss in section 2.1, this inconsistency present in the notion of {\it pure state} is then extended to the notion of {\it mixture} which is explicitly grounded on the former.

\subsection{The Inconsistency of Separability}\label{section-separability}

Something similar can be observed with regards to the notion of {\it separability}, linked in the orthodox literature to the factorizability of states and interpreted as the separation of systems into sub-systems. Regardless of the fact that the notion of quantum state is already ill defined, the notion of separability is just incompatible with the vectorial formalism of SQM. The notion of separability is grounded on the modern metaphysical representation provided by classical physics according to which physical reality is composed of independent separated individual entities which exist within space and time. According to this supposedly ``commonsensical'' picture, a system can be understood in terms of its parts and the knowledge of these parts implies the knowledge of the whole system. This is of course a direct consequence of the underlying Boolean logic that we find in classical mechanics. Indeed, as it is well known, the propositions derived from classical mechanics can be arranged in a Boolean lattice (see for a detailed discussion \cite{deRondeFreytesDomenech18}). According to classical logic, and following set theory, the {\it sum} or {\it union} of the elements of a system imply its complete characterization as a whole. However, as it is also well known since the famous paper by Birkhoof and von Neumann \cite{BvN36}, the underlying logic of QM is not Boolean, it is not {\it distributive}. Thus, the basic classical way of reasoning about systems becomes precluded right from the start. This is an obvious consequence of the fact that vectorial spaces do not relate between each other following the same rules as the elements of a set through {\it union} and {\it conjunction}. In the quantum case the equivalent to the {\it union} of two vectors is not the {\it sum} of the individual vectors considered as lines, but instead what they are capable to {\it generate} in terms of subspaces. If we consider the {\it sum} of two vectors what we obtain is not the sum of two lines but the whole plane. This shows how the basic rules of classical reasoning break down right from the start in the quantum formalism.

As it could have been easily foreseen, the artificial {\it ad hoc} introduction of a set of logical relations completely alien and even incompatible with the mathematical formalism of the theory could only lead to confusions, contradictions and pseudo-problems. Sadly enough, this is exactly what happened with the introduction of the notion of separability in the context of quantum entanglement. As we just explained, the \emph{union} of two vectors was inadequately understood as a \emph{sum} when, in fact, it is a \emph{generation}. Analogously, the \emph{projection} of a subspace was incorrectly interpreted as a {\it separation} of the whole set and the choice of a subset of elements, when, as a matter of fact, its correct interpretation is that of {\it shadow} \cite{deRondeMassri23} (figure 1). 

\begin{center}
\includegraphics[scale=.3]{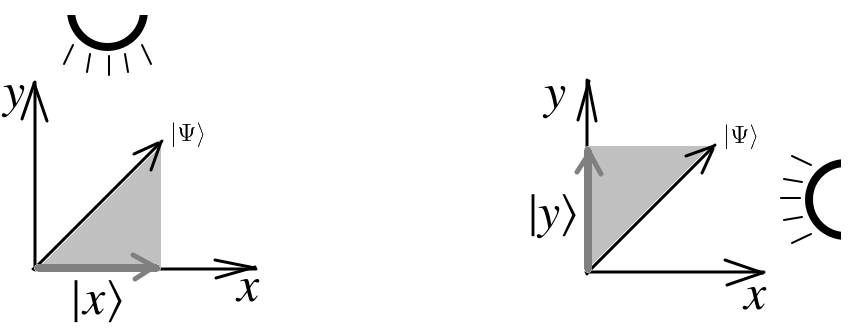} 
\captionof{figure}{The {\it shadow} of $|\Psi\rangle$ in the $x$-axis, $|x\rangle$, and in the $y$-axis, $|y\rangle$.}
\end{center}

Combining inconsistent notions --supplemented with a mythical narrative about microscopic particles-- we ended up with a meaningless definition in terms of ``the separability of pure states'' obscurely explained in terms of ``the non-local correlation between quantum particles''. Unfortunately, the contradictions do not stop at this ground level. They are also extended to the consideration of the measure of entanglement, where it is also easy to find inconsistencies and pseudo-problems within the two main lines of research present within the literature. Let us address these approaches in some detail.

\section{The Contemporary (Inconsistent) Measures of Entanglement(s)}

As we have made clear in the previous section, the reference to particles, purity and separability are non-starters for any rational analysis or research program which attempts to provide consistent definitions. These notions render impossible the construction of a bridge between the mathematical formalism, the physical concepts and their operational content. In short, they are completely fuzzy notions, full of inconsistencies and incoherencies. Unfortunately, these notions continue to ground the main approaches that appear in the literature about quantum entanglement. Indeed, we find two main lines of research which attempt to account for the measure and quantification of entanglement in terms of purity and separability. Firstly, we find an abstract geometrical approach where separability is considered in relation to entropy, and secondly, a pragmatic-operational attempt to account for the measure of entanglement in terms of Bell inequalities and classical communication. Both of these approaches make use of the fictional narrative of quantum particles and ground themselves in the inconsistencies we have already discussed leading, in turn, to deep problems. Let us discuss them in some detail.

\subsection{The Entropic Measure of {\it Geometrical Entanglement}}

One of the consequences of the vectorial formulation imposed by Bohr, Dirac and von Neumann during the early 1930s is that it becomes impossible to consider entanglement phenomena. When considering the orthodox definition of entanglement in terms of the separability of systems and subsystems we reach a limit of representation within the vectorial formulation. This is one of the main reasons behind the reconsideration during the 1990s of the matrix formalism originally proposed by Heisenberg. As explained by Hall: 
\begin{quote}
``[...] the state of a quantum system to be described by a unit vector in the corresponding Hilbert space, or more properly, an equivalence class of unit vectors under the equivalence relation $\psi\sim e^{\theta}\psi$. We will see in this section that this notion of the state of a quantum system is too limited. We will introduce a more general notion of the state of a system, described by a density matrix. The special case in which the system can be described by a unit vector will be called a pure state.''  \cite[p. 419]{Hall13} \end{quote}

Within the geometric approach the geometric representation of the state space of a two-level quantum system is given by the {\it Bloch sphere}. In this abstract representation, each point on the surface of the sphere uniquely corresponds to an abstract pure state (definition 1.2) of the Hilbert space of complex dimension 2, which characterizes a two-level quantum system, or in quantum computing terminology, a qubit. Each pair of diametrically opposed points on the Bloch sphere corresponds to two orthonormal states in the Hilbert space. For instance, the point with Cartesian coordinates $(0,0,1)$ corresponds to the state $|0\rangle$ which is orthonormal to the state $|1\rangle$ that corresponds to the opposite point $(0,0,-1)$. In general, any point on the Bloch sphere is a quantum state or qubit that is expressed in the following way: 
\[
|\psi\rangle = \cos(\theta/2)|0\rangle + e^{i\phi}\sin(\theta/2)|1\rangle
\]
Where $\theta$ and $\phi$ are real numbers such that $0 \leq\theta<\pi$ and $0 \leq\phi<2\pi$. Furthermore, the \emph{Bloch ball} whose boundary is the Bloch sphere represents the space of density matrices. In short, this geometric representation attempts to provide a visualization of quantum states and operations in quantum computing. For example, {\it quantum gates} can be represented as rotations on the Bloch sphere providing an intuitive way to understand how quantum states are transformed (see \cite{geo-q-s}). In the case of quantum systems with more than two levels, called {\it qudits}, the state space becomes a higher-dimensional complex projective space. Visualizing this space directly is challenging due to our inherent difficulty in picturing high-dimensional spaces. Within the literature the application of the Bloch sphere representation has motivated to seek similar geometric approaches for higher-dimensional state spaces. One such approach is to generalize the Bloch sphere to higher-dimensional varieties. In this representation, a d-level quantum system is associated with a $(2d-1)$-dimensional real vector space. While this approach does not provide an intuitive picture, it does offer a way to visualize and analyze multi-level quantum systems using geometric concepts which can be useful when attempting to understand complex quantum operations and phenomena in higher dimensions. It is important to understand that the geometric approach to SQM emphasizes the study of the shape and properties of the state space in an abstract setting --avoiding a coordinate representation. The mathematical approach is called {\it coordinate-free} because it focuses on properties that are independent of the choice of coordinates. Quantum states are then viewed as points in a complex projective space, and quantum operations as transformations of this space. The main attempt is to represent the properties of quantum systems, such as separability, in terms of the geometry of the mathematical space. For example, two quantum states are orthogonal if and only if their corresponding points in projective space are separated by a distance of $\sqrt{2}$. Similarly, a quantum operation that preserves the orthogonality between states corresponds to an isometry (distance-preserving transformation) of the projective space. 

However, the elimination of reference frames and coordinate systems within this highly abstract mathematical approach to physical representation leads also to the elimination of the operational content of the theory. This becomes explicit when considering the notions of pure and mixed states. In fact, as pointed out above, there is an essential inconsistency in the literature between the purely abstract vectorial definition (definition 1.2) and the basis-dependent operational definition (definition 1.1). This inconsistency also leads to serious difficulties within the geometrical approach. Consider the set $B(H)$ of bounded operators (abstract matrices) in a finite dimensional Hilbert space $H$ (abstract vector space). While an {\it abstract pure state} is defined as a rank one hermitian operator (i.e., $\rho\in B(H)$ such that $\rho=\rho^\dag$ and $\text{rk}(\rho)=1$), an {\it abstract mixed state} is defined as a convex combination of abstract pure states (figure 2). This is an abstract density matrix which has rank bigger than 1 (abstract pure states have rank equal to 1). 
\begin{center}
\includegraphics[scale=.5]{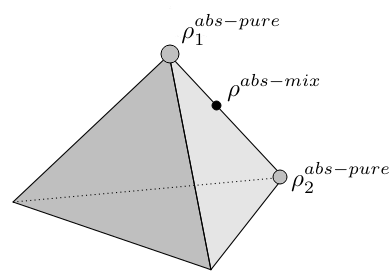}
\captionof{figure}{Abstract pure and mixed density matrices.}
\end{center}
Now, unlike pure states, which are considered as an accurate description of quantum entities, abstract mixed states are interpreted in terms of  ignorance regarding which {\it is}, in fact, the underlying actual pure state. Specifically, if an abstract mixed state is given by the following convex combination,
\[
\rho^{abs-mix} = \sum_{i=1}^n p_i \rho_{i}^{abs-pure},
\]
then it is interpreted that the state of the system is in one of the pure states $\rho_{i}^{abs-pure}$ with probability $p_i$.
According to the orthodox interpretation, this notion is purely epistemic, it does not provide an accurate account of which is the quantum state but, like in classical probability, quantifies our ignorance regarding the underlying state of affairs described in terms of pure states. However, it should be noticed that, as we discussed above, an abstract pure state --since it lacks any operational content-- does not really provide a description of a physical state of affairs. Thus, it becomes unclear what this epistemic interpretation of mixtures is really about. In general, it is always possible to express $\rho^{abs-mix}$ as a convex combination of different abstract pure states. It is in fact possible to write $\rho^{abs-mix}$ in different ways, 
\begin{align*}
\rho^{abs-mix} &= p_1{\rho}_{1}^{abs-pure}+p_2{\rho}_{2}^{abs-pure}+p_3{\rho}_{3}^{abs-pure}\\
&= q_1{\rho}_{1}^{abs-pure}+q_2{\delta}_{2}^{abs-pure}+q_3{\delta}_{3}^{abs-pure}
\end{align*}
This account leads to contradictions: while on one hand $\rho^{abs-mix}$ has probability $p_1$ of being $\rho_{1}^{abs-pure}$, on the other hand it has probability $q_1$ of being the same state $\rho_{1}^{abs-pure}$. The same mixture has two different probabilties of being in the same pure state. Even worse, one might have two different accounts of the same mixture which do not share any common pure state. As an exmaple: 
\begin{align*}
\rho^{abs-mix} &= p_1{\rho}_{1}^{abs-pure}+p_2{\rho}_{2}^{abs-pure}+p_3{\rho}_{3}^{abs-pure}\\
&= r_1{\gamma}_{1}^{abs-pure}+r_2{\gamma}_{2}^{abs-pure}+r_3{\gamma}_{3}^{abs-pure}
\end{align*}
While in the first case we obtain that the mixture should be described in terms of ${\rho}_{1}^{abs-pure}$, ${\rho}_{2}^{abs-pure}$ or ${\rho}_{3}^{abs-pure}$ in the second case we obtain that the same mixture is not  any of these ${\rho}_{i}^{abs-pure}$ states but, instead, either ${\gamma}_{1}^{abs-pure}$, ${\gamma}_{2}^{abs-pure}$ or ${\gamma}_{3}^{abs-pure}$. Clearly, according to the epistemic interpretation of mixtures, the underlying state cannot be at the same time given by one of the ${\rho}_{i}^{abs-pure}$ and none of the ${\rho}_{i}^{abs-pure}$.  

\smallskip

Now, according to the orthodox literature, the essential relevance of the geometrical approach is that, in contradistinction to the vectorial formulation, one can define an abstract separable pure state (or abstract product state). Assume that we have a specific (abstract) factorization $H=H_1\otimes H_2$. Then,  an {\it abstract separable pure state}  is an abstract pure state in $H$ which is a product  between an abstract pure state in $H_1$ and a abstract pure state in $H_2$, 
\[
\rho^{abs-pure}  = \rho^{abs-pure}_1\otimes \rho^{abs-pure}_2.
\]
As in the orthodox account, the abstract separable pure state is then interpreted as a state consisting of two independent subsystems. An {\it abstract separable mixed state} is a convex combination of separable pure states: 
\[
\rho^{abs-mix} = \sum_{i=1}^n p_i \rho_{1i}^{abs-pure}\otimes \rho_{2i}^{abs-pure}
\]
The mixed state $\rho^{abs-mix} $ is actually in the bipartite state $\rho_{1i}^{abs-pure}\otimes \rho_{2i}^{abs-pure}$ with probability $p_i$. However, the same problem arises as before. The same separable mixed state can be interpreted in a contradictory fashion, for example, as being in one of the states  $\gamma_{1i}^{abs-pure}\otimes \gamma_{2i}^{abs-pure}$, in which case, it is in none of the states $\rho_{1i}^{abs-pure}\otimes \rho_{2i}^{abs-pure}$.  

The contradiction also appears in the inconsistent combination of notions, for an {\it entangled state} is defined in this scheme as one which is not a {\it separable mixed state}. But while the notion of {\it abstract pure state} is ontological, the notion of {\it abstract mixed state} is epistemic. This implies that while the notion of {\it abstract separable pure state} is ontological, the notion of {\it abstract separable mixed state} is epistemic. Now, when extending this line of reasoning in the geometric approach and defining entanglement in terms of the epistemic notion of mixed separable state, we end up characterizing entanglement in terms of what is not something (the abstract separable mixed state) that we do not know what it is (which pure state). We end up defining entanglement as the negation of an epistemic notion, namely, as the negation of the ignorance of something that it is not.

This shows that the basic notions and definitions applied within the approach are ill defined. But the problems do not stop here. To the already problematic notions we have described we need to add the strange introduction of the notion of  {\it entropy} in order to account for the measure of entanglement. The term entropy pertains originally to classical thermodynamics. Then, however, it was exported to statistical mechanics, information theory and other lines of research, where the notion was ambiguously transformed. Entropy became then a scientific concept associated with a  state of disorder, but also --although without much clarity-- with randomness and uncertainty. One of those unclear transplants of the notion of entropy, perhaps the most recent one, determined its use for the study of entanglement.
\begin{center}
\includegraphics[scale = .7]{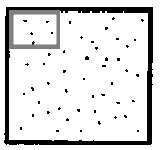}
\captionof{figure}{The entropy of the subsystem \\is less than the entropy of the system.}
\end{center} 
In QM the notion of entropy that it is used is the {\it von Neumann entropy},
\[
S(\rho)=-\text{Tr}(\rho\log(\rho)).
\]
Notice that this definition is invariant and works for operational and abstract states. 
One of the key properties of von Neumann entropy is the following equivalence:
\[
\rho\text{ is pure}\iff S(\rho)=0.
\]
But what is the physical interpretation of this notion? In relation to purity the orthodox interpretation is that a pure state has {\it maximal certainty} and consequently it is also totally ordered. But as we have already remarked, this definition of purity (definition 1.2) is independent of the basis (any rank one matrix has maximal certainty) and stands in contraposition with the definition of operational-purity according to which the only matrix which provides maximal certainty is the diagonal matrix $(1,0,\dots,0)$. In other words, while the state $|0\rangle$ is operationally pure (it has complete certainty), the same state in a superposition $(|\uparrow\rangle+|\downarrow\rangle)/\sqrt{2}$ is not  operationally-pure because there is complete uncertainty of the outcomes that will be obtained. However, when considering abstract purity, both have zero entropy.

Another  property of von Neumann entropy that is used as a criterion to study separability is the additivity for product states, $S(\rho_1\otimes\rho_2)=S(\rho_1)+S(\rho_2)$. 
In order to interpret this property, we must assume that $\rho$ represents a bipartite system and its reduced states are subsystems. For a separable state, the criterion says that the \emph{uncertainty} of the system is greater than the \emph{uncertainty} of its subsystems (as in the case of classical thermodynamics shown in figure 3). But as shown in \ref{section-separability} this notion is ill defined and the notion of entropy is inadequate in this framework first of all, because it is relating certainty with purity as before. Furthermore, the abstract definition of non-separability (or entanglement) is conceptually misleading and operationally impractical: there is no way of saying if a state is entangled or not. In fact, this is a problem classed as NP-hard, namely, a problem for which no efficient algorithm is currently known that can solve all instances of the problem in polynomial time (see \cite{Gharibian}).\footnote{The concept of NP-hardness is crucial in understanding the difficulty of various computational problems. Many real-world problems from different fields, such as optimization, scheduling, and graph theory, have been shown to be NP-hard.} This implies that, unless P (the class of problems that can be solved in polynomial time) equals NP, there's no known way to solve NP-hard problems efficiently for all possible inputs. Thus, we rapidly reach a practical impossibility for applying this property to find out if a state is separated or not. 
This has led to lower expectations, to a certain resignation, and to reframe the study of separability in terms of a much weaker property called \emph{entanglement witness}. 
\begin{center}
\includegraphics[scale=.7]{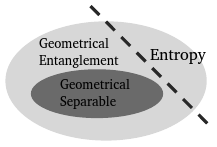} 
\captionsetup{justification=centering}
\captionof{figure}{If the state is on the left side of the diagonal\\ it is impossible to know if it is entangled or not.}
\end{center}
Entanglement witness is a function (linear or nonlinear, see \cite{Guhne}) that is incapable to determine in all cases whether a state is entangled or not (figure 4).\footnote{An interesting mathematical property that can be used to construct an entanglement witness is the \emph{majorization criterion} which relates the eigenvalues of the state with the eigenvalues of its reduced states. Specifically,  if the eigenvalues of $\rho$ are $p=(p_1\ge p_2\ge \dots)$ and the eigenvalues of its reduced matrix $\rho_A=\text{Tr}_B(\rho)$ are $q=(q_1\ge q_2\ge\dots)$,  then for any separable state $\rho$ we must have $q\preceq p$. In particular, this criterion implies that if $\rho$ is separable, then the entropy of $\rho$ is greater than the entropy of $\rho_A$ (same for $\rho_B$) (see \cite[Th. 1]{Horodecki} for more information).} 

\subsection{The Operational Measure of {\it Communicational Entanglement}}

In the orthodox literature, apart from its definition in terms of non-separability, quantum entanglement has been also characterized in terms of ``non-locality'' --as a type of ``a non-classical correlation''. Historically, this has been directly linked to the re-consideration of entanglement by a particle physicist working at CERN during the mid 1960s. In secrecy, using only weekends and after hours, John Bell would go back to the forbiden EPR paper in order to come up with a testable experiment that would confirm his expectation to describe quantum phenomena in classical terms. In order to constraint quantum phenomena Bell would re-discover a series of inequalities that Boole had already constructed a century before when analyzing the constrains of classical experience in terms of classical probability (see \cite{Pitowsky94}). As it is well known in the philosophical circles, Boole-Bell inequalities are a statistical statement about classical probability, not about QM. This means that even though Boole-Bell inequalities constrain the statistical correlations found within classical probability theory, they say nothing about the theory of quanta. Thus, testing these inequalities can only expose the impossibility to model experimental data in classical terms. The violation of Bell inequalities does not logically imply the existence of quantum phenomena. And, conversely, it's non-violation does not imply the reference to classicality. Unfortunately, it is exactly the contrary which has become the orthodox interpretation of Bell inequalities. 

The reasons behind this inconsistent line of reasoning are linked to the Bohrian narrative dogmatically established within the Standard account of the theory according to which ``QM describes a microscopic realm of quantum particles'', and also, that ``there is a {\it limit} between the quantum and classical descriptions'' --something which Bohr dogmatically imposed through his {\it correspondence principle}. It is these unjustified dogmas which provide an implicit justification for the claim that ``what is not classical must be necessarily regarded as being quantum'', and vice-versa,  ``what is not quantum must be classical''. It is in this orthodox context that Bell's statement about classical probability --derived a century earlier by Boole, when QM did not even exist as theory-- has come to be regarded as an essential component of the theory of quanta expressing its non-local behaviour. For example, in a recent review we can read the following:  
\begin{quotation}
\noindent {\small ``Bell's theorem has deeply influenced our perception and understanding of physics, and arguably ranks among the most profound scientific discoveries ever made. With the advent of quantum information science, a considerable interest has been devoted to Bell's theorem. In particular, a wide range of concepts and technical tools have been developed for describing and studying the nonlocality of quantum theory.'' \cite{Brunner}}
\end{quotation}
As revealed by the authors of the just mentioned review: ``In the last two decades, Bell's theorem has been a central theme of research from a variety of perspectives, mainly motivated by quantum information science, where the nonlocality of quantum theory underpins many of the advantages afforded by a quantum processing of information.'' And in he same review \cite{Brunner} we find also the widespread reference to particles: ``a typical `Bell experiment', two systems which may have previously interacted --for instance they may have been produced by a common source-- are now spatially separated and are each measured by one of two distant observers, Alice and Bob.''  Regardless of the reference to particles, according to many, this leads to a purely operational, non-ontological approach to quantum entanglement which is then studied in terms of many different conditions of information transfer such as {\it Local Operations and Classical Communication} (LOCC) or {\it Local Operations and Shared Randomness} (LOSR). In this case quantum correlations are conceived as a resource. 

Let us review briefly the theory of \emph{quantum instruments} from  \cite{Chitambar}. A (discrete) quantum instrument is a family of completely positive (CP) maps $\{\mathcal{E}_j\}$ such that $\sum_j\mathcal{E}_j$ is trace-preserving.  When it is applied to the state $\rho$, $\mathcal{E}_j(\rho)$ represents the postmeasurement  state associated with the outcome $j$, which occurs with probability $\text{Tr}(\mathcal{E}_j(\rho))$. For an $n$-partite quantum system, the underlying state space is $H := H_1 \otimes \dots\otimes H_n$  with $H_k$ being the reduced state space of party $k$. An instrument is called \emph{one-way local} with respect to party $k$ if each of its CP maps has the form $\mathcal{T}_1\otimes\dots\otimes\mathcal{T}_{k-1}\otimes\mathcal{E}_k \otimes\mathcal{T}_{k+1}\otimes\dots\otimes\mathcal{T}_n$,  where $\mathcal{E}_k$ is a CP map on $B(H_k)$, and for each $j\neq k$, $\mathcal{T}_j$
is some  trace-preserving completely positive (TCP) map. This one-way local operation consists of party $k$ applying an instrument $\{\mathcal{E}_j\}$, broadcasting the classical outcome to all other parties, and party $j$ applying TCP map $\mathcal{T}_j$ after receiving this information. Operationally, $\text{LOCC}_r$ is the set of all instruments that can be implemented by some $r$-round LOCC protocol. Here, one round of communication involves one party communicating to all the others, and the sequence of communicating parties can depend on the intermediate measurement outcomes. The set of instruments that can be implemented by some finite round protocol is then $\text{LOCC}_{\mathbb{N}}$. On the other hand, so-called infinite round protocols, or those having an unbounded number of non-trivial communication rounds, correspond to instruments in  $\text{LOCC} \setminus \text{LOCC}_\mathbb{N}$.  The full set of LOCC then consists of both bounded-round protocols as well as the unbounded ones. The set $\overline{\text{LOCC}_\mathbb{N}}$ is the topological closure of $\text{LOCC}_\mathbb{N}$, and it is equal to $\overline{\text{LOCC}}$. To complete the picture, let us provide the definition related to separable (SEP) instruments. A multipartite state $\rho$ is called (fully) separable if it can be expressed as a convex combination of product states with respect to a partition. Operationally, this class is more powerful than LOCC, yet it is still more restrictive than the most general quantum operations. At the same time, it admits a simpler mathematical characterization than LOCC, and this can be used to derive many limitations on LOCC such as entanglement distillation and state discrimination. The classes of LOCC and SEP are related by the following chain of inclusions,
\[
\text{LOCC}_1\subsetneq\dots \text{LOCC}_r\subsetneq \text{LOCC}_\mathbb{N}\subsetneq \text{LOCC} \subsetneq  \overline{\text{LOCC}}\subsetneq \text{SEP}.
\]

It is of course not strange that this path has led to many pseudo-problems which ended up exposing the shaky foundations of the whole program. As resumed by Bokulich and Jaeger: 
\begin{quotation}
\noindent {\small``There are [...] limitations to using a violation of Bell's inequality as a general measure of entanglement. First, there are Bell-type inequalities whose largest violation is given by a non-maximally entangled state (Ac\'in et al. 2002), so entanglement and non-locality do not always vary monotonically. More troublingly, however, Reinhard Werner (1989) showed that there are some mixed states (now referred to as Werner states) that, though entangled, do not violate Bell's inequality at all; that is, there can be entanglement without non-locality. In an interesting twist, Sandu Popescu (1995) has shown that even with these local Werner states one can perform a non-ideal measurement (or series of ideal measurements) that `distills' a non-local entanglement from the initially local state. In yet a further twist, the Horodecki family (1998) subsequently showed that not all entanglement can be distilled in this way --there are some entangled states that are `bound'. These bound entangled states are ones that satisfy the Bell inequalities (i.e., they are local) and cannot have maximally entangled states violating Bell's inequalities extracted from them by means of local operations. Not only can one have entanglement without non-locality, but also, as Bennett et al. (1999) have shown, one can have a kind of `non-locality without entanglement'. There are systems that exhibit a type of non-local behavior even though entanglement is used neither in the preparation of the states nor in the joint measurement that discriminates the states (see also Niset and Cerf (2006))'' \cite[pp. xvii-xviii]{BokulichJaeger10} }\end{quotation}

The idea that Boole-Bell inequalities talk about QM is an error which, like many others, regardless of rational argumentation, has become naturalized in the foundational literature creating a maze of many fragmented fields characterized by inconsistency and vagueness. Of course, there is a way to expose this error, even when considering the inconsistent notions applied within the field. That is, even under the conditions imposed by a wrong interpretation of Bell inequalities, the distinction they provide is essentially useless even from an instrumentalist viewpoint. Take for example the state 
\[
p|\phi_+\rangle\langle \phi_+| + (1-p)\frac{\mathbf{1}}{4}
\]
which is separable for $p\le 1/3$ and entangled otherwise and does not violate any Bell inequality.
(For more examples and a more technical discussion, see \cite{Brunner}.) The problem is exposed graphically in figure 5.
\begin{center}
\includegraphics[scale=.8]{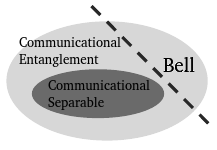}
\captionof{figure}{Bell inequalities and the impossibility to\\distinguish separable from non-separable states}
\end{center}
Even when accepting the (inconsistent) orthodox definition of separability, Bell inequalities are simply incapable to distinguish between separable and non-separable states.

\subsection{An Entangled Map of Madness?}

The conclusion we can draw after our critical analysis is that there appears to be something deeply flawed in the mainstream contemporary approach to quantum entanglement. We lack a meaningful definition, we have no understanding of what occurs in entangled phenomena, and, moreover, we possess no consistent way of measuring and determining the existence of entanglement. Furthermore, the literature on the subject exhibits an extreme fragmentation where different essentially inconsistent definitions, grounded themselves on inconsistent dogmas, coexist in different sub-domains of research. Indeed, from the orthodox widespread definition of entanglement in terms of the non-separability of pure states we find two different developments which are not equivalent. On the one hand, a geometrical definition in terms of an extremely unclear notion of entropy, and on the other, a communicational definition in terms of the violation of Bell type inequalities which, in turn, is also fragmented into many different sub-definitions, depending on the specific conditions of communicability (e.g., LOCC, LOSR, etc.). It is at this point that we might characterize this communion between inconsistency and fragmentation --borrowing Adan Cabello's characterization of the interpretational debate about QM \cite{Cabello17}-- as an entangled ``map of madness''. 

Rethinking the situation we can conclude, perhaps, that what is fundamentally wrong are the assumptions we are taking for granted without criticism, which were determined through the establishment of SQM almost a century ago. These orthodox presuppositions were responsible for the destruction of the invariance of the theory, of its formal and operational consistency and --even-- of the concept of (quantum) {\it state}. They have also introduced perspectival relativism at the core of the theory and propagated a series of vaguely defined or directly inconsistent notions such as {\it purity} and {\it separability}. But the truth is that there is a way out of this labyrinth, an exit of this maze we can find by simply following the thread of operational invariance.

\section{The Thread of Operational Invariance}

In principle, at least in formal terms, it is operational invariance which defines what has to be considered as the \emph{same} in a physical theory. The essential role played by operational invariance consists in that it allows to consider experience from different reference frames in a consistent manner as referring to the same state of affairs. We might talk here of an objective relativism since, as Einstein constantly remarked, this makes possible to address the existence of a state of affairs detached from particular observations and reference frames. Given a mathematical formalism of a theory which has an invariant transformation, it does not really matter which particular reference frame we might choose to describe the state of affairs simply because there will be a consistent translation between any of them. We can go from one representation to another without loosing consistency and coherency regarding what is considered to be {\it the same} state.\footnote{Of course, thsi does not mean that the properties considered from different reference frames will possess the same values (e.g., position and velocity in classical mechanics), it means there will be a consistent translation of these properties from one reference frame to the other.} This is surely realized in Newtonian mechanics, in Maxwell’s electromagnetism, and also --via the Lorentz' transformations-- in relativity theory. But is it possible to find invariance in QM? Contrary to what is usually believed, the answer is yes, and right from the start, in Heisenberg’s matrix mechanics. 

During the early 1920s, Heisenberg had followed Bohr’s guide, focusing on the question of describing the trajectories of electrons inside the atom. But the critical reaction of Wolfgang Pauli and Arnold Sommerfeld led him in 1924 to take a different path. So, instead of trying to describe trajectories of unseen, presupposed, corpuscles, Heisenberg reframed the problem in terms of observable quantities. As explained by Jaan Hilgevoord and Joos Uffink \cite{HilgevoordUffink01}: “His leading idea was that only those quantities that are in principle observable should play a role in the theory, and that all attempts to form a picture of what goes on inside the atom should be avoided. In atomic physics the observational data were obtained from spectroscopy and associated with atomic transitions. Thus, Heisenberg was led to consider the --transition quantities-- as the basic ingredients of the theory.” One year later, he would present his groundbreaking results in the following manner \cite{Heis25}: “In this paper an attempt will be made to obtain bases for a quantum-theoretical mechanics based exclusively on relations between quantities observable in principle.” Emancipating himself completely from the atomist discourse, Heisenberg was able to create a completely new mathematical formalism. As he would recall in his autobiography:
\begin{quotation}
\noindent {\small ``In the summer term of 1925, when I resumed my research work at the University of G\"ottingen --since July 1924 I had been {\it Privatdozent} at that university-- I made a first attempt to guess what formulae would enable one to express the line intensities of the hydrogen spectrum, using more or less the same methods that had proved so fruitful in my work with Kramers in Copenhagen. This attempt lead me to a dead end --I found myself in an impenetrable morass of complicated mathematical equations, with no way out. But the work helped to convince me of one thing: that one ought to ignore the problem of electron orbits inside the atom, and treat the frequencies and amplitudes associated with the line intensities as perfectly good substitutes. In any case, these magnitudes could be observed directly, and as my friend Otto had pointed out when expounding on Einstein's theory during our bicycle tour round Lake Walchensee, physicists must consider none but observable magnitudes when trying to solve the atomic puzzle.'' \cite[p. 60]{Heis71}}
\end{quotation}

Heisenberg was capable of developing matrix mechanics following two ideas: first, to leave behind the classical notion of particle-trajectory, as it did not seem required by QM —and it rather appeared as a classical habit that was inadvertently coloring a priori the approach to the understanding of the new theory—, and second, to take as a methodological standpoint Ernst Mach’s positivist idea according to which a theory should only make reference to what is actually observed in the lab. And what was actually observed was well known to any experimentalist: a spectrum of line intensities. This is, in fact, what was described by the tables of data that Heisenberg attempted to mathematically model and that finally led him —with the help of Max Born and Pascual Jordan— to the development of the first consistent mathematical formulation of the theory of quanta. Let us stop to take note once again of some of the conditions that were fundamental for the development of the quantum formalism. First, Heisenberg’s abandonment of Bohr’s atomist narrative and research program which focused in the description of unobservable trajectories of presupposed yet irrepresentable quantum particles. Second, the consideration of Mach’s observability principle as a methodological standpoint that --even if Heisenberg didn't fully embraced the positivist credo-- allowed him to find a starting point unburdened of those classical presuppositions. That methodological standpoint led him finally to the replacement of Bohr’s fictional trajectories of irrepresentable electrons by the consideration of the intensive quantities appearing in line spectra that were actually observed in the lab. And these quantities, once detached from a supposedly necessary reduction to atomic elements, were what the formalism was indicating as invariant. Radically new, and of fundamental importance to produce a consistent and invariant quantum formalism, was this idea that we should accept intensive values as basic, as perfectly good “substitutes”, that are in no need whatsoever to be reconduced to binary values. Intensities appeared as basic and sufficient. But Heisenberg's intuition, according to which we should take as mainly significant the intensive patterns, was mostly discarded. The replacement of intensive patterns by the unilateral focus on single outcomes, by the idea that single outcomes are what is most meaningful, took place especially in Dirac's work, and as a consequence of a dogmatic presupposition with neither theoretical nor experimental justification: we must refer to particles, and those single outcomes must be the expression of specific particles. This idea according to which single binary outcomes are what must be explained, what the theory talks about, entailed leaving aside the intensities that appeared as invariant in the formalism. And, as invariance was destroyed, there appear the need to embrace relativism as acceptable, and to produce a necessary projection postulate, since the (intensive) formalism was of course incapable of predicting single measurement outcomes.

In any case, a physical theory is not only an invariant mathematical formalism. Equally fundamental is the development of a conceptual representation that allows to qualitatively understand what is physically real according to the formalism, and to give meaning to the observations that are predicted by the formalism. Together with formal invariance, conceptual objectivity is a fundamental aspect of any physical theory. But this conceptual representation cannot be an arbitrary addition. An objective conceptual representation has to be developed in strict accordance with the conditions established by the formalism. Specifically, it should be grounded on the {\it moment of unity} produced formally by the invariant elements of the formalism. It must construct an objective conceptual representation of what appears as invariant, producing the physical concepts that correspond to these invariant elements. If, as we said, what appears as invariant are the intensive quantities, and if this invariance is lost when we attempt to redirect intensities to binary values, we should start by producing a concept that is originally intensive, which is sufficient, which does not entail the redirection to other elements that could be understood in a binary manner. In this respect, we propose the concept of \emph{power of action}, or \emph{intensive power}, which represents those invariant intensive quantities present in the quantum formalism (turning unnecessary the need of adding the projection postulate). The intensive value of each power is termed its \emph{intensity} or \emph{potentia}. In fact, if we think about it, the reference to a physical reality of an intensive nature was already present right from the start in the way Max Planck formulated his original discovery: the quantum of \emph{action}. Action surely is no particle; action represents a reality perhaps more intuitively conceived in intensive terms. The reference is also found in {\it configuration space} which is nothing but a space which encapsulates {\it degrees of action}. In this representation, QM talks about ``action'', about powers of action, quantitatively defined by their intensity or potentia. This representation entails a rejection of the atomistic image of the world, that forces us constantly to take the probabilistic values as a diminished, insufficient representation, as evidently not the --real thing’, and it also entails that it is not necessary to redirect those intensive quantities in each case to binary values in order to determine a state of affairs. Let us add that the notion of a physical element that is in itself intensive implies a different understanding of observation. For instance, in the particular experimental situations where we obtain a single outcome at a time, it is only possible to obtain a measure of the physical element considered within the theory through repetition. This means that, in those situations, a single outcome is not what is mainly meaningful, but, on the contrary, an insufficient information, a minimal and partial measure. This is an example of how the understanding of observation is conditioned by theoretical presupositions --contrary to the contemporary widespread empiricist understanding according to which theories begin with uncontaminated observations.

When we stick to the intensive values of action, we are able in fact to refer to a state of affairs that is independent of the particular representation in a reference frame (or basis), escaping thus the relativism with which most accounts of QM have contented themselves. In contraposition to an Actual State of Affairs, defined classically in terms of a set of true definite valued binary properties, we propose to relate the reference of QM to an Intensive State of Affairs (ISA) --also called elsewhere Potential State of Affairs (PSA) \cite{deRondeMassri21a}. What has been demonstrated is that by considering an {\it intensive}, rather than {\it binary}, state of affairs, it is possible to restore a consistent global valuation for all projection operators independently of the basis. Let us recall some results from \cite{deRondeMassri21a}. While a \emph{Global Binary Valuation} (GBV) is a function from a graph to the set $\{0,1\}$, a \emph{Global Intensive Valuation} (GIV) is a function from a graph to the closed interval $[0,1]$. We term projection operators as {\it intensive powers}.\footnote{For a detailed introduction, analysis and discussion of the notion of `intensive power' we refer the interested reader to \cite{deRonde16}, and more specifically, \cite[Sect. 8]{deRondeMassri21a} and \cite[Sect. 3]{deRondeMassri19a}.} Let $H$ be a Hilbert space and let $\mathcal{G}=\mathcal{G}(H)$ be the set of observables. We give to $\mathcal{G}$ a graph structure by assigning an edge between observables $P$ and $Q$ if and only if $[P,Q]=0$. We call this graph, \emph{the graph of powers}. Among all global intensive valuations we are interested in the particular class of ISA.
\begin{definition}
Let $H$ be a Hilbert space. An \emph{Intensive State of Affairs} is a global intensive valuation
$\Psi:\mathcal{G}(H)\to[0,1]$ from the graph of powers $\mathcal{G}(H)$
such that $\Psi(I)=1$ and 
\[
\Psi(\sum_{i=1}^{\infty} P_i)=
\sum_{i=1}^\infty \Psi(P_i)\]
for any piecewise orthogonal projections $\{P_i\}_{i=1}^{\infty}$.
The numbers $\Psi(P) \in [0,1]$, are called {\it intensities} or {\it potentia}
and the nodes $P$ are called \emph{ powers}.
Hence, an ISA assigns a potentia to each power.
\end{definition}
Intuitively, we can picture an ISA
as a table,
\[
\Psi:\mathcal{G}(H)\rightarrow[0,1],\quad
\Psi:
\left\{
\begin{array}{rcl}
P_1 &\rightarrow &p_1\\
P_2 &\rightarrow &p_2\\
P_3 &\rightarrow &p_3\\
  &\vdots&
\end{array}
\right.
\]

\begin{theorem}
Let $H$ be a separable Hilbert space, $\dim(H)>2$ and let $\mathcal{G}$ be the graph of powers with the commuting relation given by QM.
\begin{itemize}
\item Any positive semi-definite self-adjoint operator 
of the trace class $\rho$ determines in a bijective way
an ISA $\Psi:\mathcal{G}\to [0,1]$. 
\item Any GIV determines univocally a set of powers that are considered as truly existent. 
\end{itemize}
\end{theorem}

\begin{proof} 
\begin{enumerate}
\item Using Born's rule, we can assign to each
observable $P\in\mathcal{G}$ the value $\mbox{Tr}(\rho P)\in[0,1]$.
Hence, we get an ISA $\Psi:\mathcal{G}\to[0,1]$.
Let us prove that this assignment is bijective. Let 
$\Psi:\mathcal{G}\to[0,1]$ be an ISA. By Gleason's theorem
there exists a unique positive semi-definite self-adjoint operator 
of the trace class $\rho$
such that $\Psi$ is given by the Born rule with
respect to $\rho$.\footnote{As remarked in \cite{WK}: ``Prior to the Bell and Kochen-Specker theorems, Gleason's theorem demonstrated that, for any quantum system of dimension at least three, the unique way to assign probabilities to the outcomes of projective measurements is via the Born rule. In particular, Gleason's theorem excludes any deterministic probability rule given by a \{0, 1\}-valued assignment of probabilities to all the self-adjoint projections on the system's Hilbert space.''}
\item Consider the function $\tau:[0,1]\to\{0,1\}$, 
where $\tau(t)=0$ if and only if $t=0$. Now, given a 
GIV $\Psi:\mathcal{G}\to[0,1]$, the map $\tau \Psi:\mathcal{G}\to\{0,1\}$
is a well-defined map. 
\end{enumerate}
\end{proof}

\begin{definition}
Let $\mathcal{G}$ be a graph. We define a \emph{context} as a complete subgraph (or aggregate) inside $\mathcal{G}$. For example, let $P_1,P_2$ be two elements of $\mathcal{G}$. Then, 
$\{P_1, P_2\}$ is a contexts if $P_1$ is related to $P_2$, $P_1\sim P_2$. Saying it differently, if there exists an edge between $P_1$ and $P_2$. In general, a collection of elements $\{P_i\}_{i\in I}\subseteq \mathcal{G}$ determine a context if $P_i\sim P_j$ for all $i,j\in I$. Equivalently, if the subgraph with nodes $\{P_i\}_{i\in I}$ is complete.  A \emph{maximal} context is a context not contained properly in another context.  If we do not indicate the opposite, when we refer to contexts we will be implying maximal contexts.
\end{definition}

For the graph of powers, the notion of context coincides with the usual one; a complete set of commuting operators. However, all projection operators can be assigned a consistent value bypassing in this way the famous Kochen-Specker theorem \cite{KS}.

\begin{theorem} 
{\sc (Intensive Non-Contextuality Theorem)} Given any Hilbert space $H$, then an ISA is possible over $H$.
\end{theorem} 
\begin{proof}  
See \cite{deRondeMassri21a}.
\end{proof}

\noindent This theorem restores the possibility of an invariant physical representation of any quantum wave function $\Psi$. Thus, contrary to the orthodox interpretation of QM in terms of systems with properties (which dogmatically impose a binary valuation), our conceptual representation of quantum physical reality is not relative to any particular context, it is global and essentially intensive. We refer the reader to \cite{deRondeMassri19a, deRondeMassri21a} for a detailed discussion and analysis. 

This understanding of the theory allows also for the restoration of causality: QM talks about intensive values of action, that evolve casually, and that can be measured with an {\it intensive certainty} without any inconsistency (as said before, when proceeding with a single outcome at a time, this can be perfectly done through repetition). There is thus nothing ``random'', ``indeterminate'' or ``uncertain'' in QM, there are no ``quantum jumps'' or ``collapses''. In short, the problems only appear when physicists take single measurement outcomes as the reference of the theory, when they attempt to discuss about single measurement outcomes instead of referring to intensities. But, as we said, that unilateral focus on single outcomes is surely not determined by the theory, and it only arises from the unjustified presupposition according to which we are referring to particles (each single measurement is taken then as the expression of a specific presupposed particle). We might say that even though QM does not talk about clicks, it can explain their appearance in detectors just in the same way as electromagnetism can explain the magnetic attraction of two rocks even though the theory makes no reference whatsoever to such material elements.




\section{Entanglement as an Operational Expression of Quantum Powers}

In order to advance from the intensive and invariant reconsideration of QM --as proposed by the logos approach-- to a meaningful understanding of entanglement, we must consider the operational concepts that are required in order to bridge the gap between the quantum formalism and what is actually observed in the lab. Let us thus review some definitions proposed in \cite{deRondeFMMassri23a}. 

A {\bf screen} with $n$ {\bf detectors} corresponds to the vector space $\mathbb{C}^n$. Choosing a basis, say $\{|1\rangle,\dots,|n\rangle\}$, is the same as choosing a specific set of $n$ detectors. A {\bf factorization} $\mathbb{C}^{i_1}\otimes\dots \otimes\mathbb{C}^{i_n}$ is the specific number $n$ of screens, where the screen number $k$ has $i_k$ places for detectors, $k=1,\dots,n$. Choosing a {\bf basis} in each factor corresponds to choosing the specific detectors; for instance $|\uparrow\rangle, |\downarrow\rangle$. After choosing 
a basis in each factor, we get a basis of the factorization $\mathbb{C}^{i_1}\otimes\dots \otimes\mathbb{C}^{i_n}$
that we denote as
\[
\{ |k_1\dots k_n\rangle \}_{1\le k_j\le i_j}.
\]
The {\bf basis element} $|k_1\dots k_n\rangle$ determines the {\bf  projector}  $|k_1\dots k_n\rangle \langle k_1\dots k_n|$ which is the formal-invariant counterpart of the objective physical concept called {\bf power of action} (or simply \textbf{power}) that produces an intensive global effect in the $k_1$ detector of the screen $1$,  in the $k_2$ detector of the screen $2$ and so on until the $k_n$ detector of the screen $n$. Let us stress the fact that this intensive effectuation does not allow an explanation in terms of particles within classical space and time. Instead, this is explained as a characteristic feature of any quantum power. In general, any given power will produce a unitary multi-screen non-local effect that has an intensive content. 

Given an ISA, $\Psi$, a factorization $\mathbb{C}^{i_1}\otimes\dots \otimes\mathbb{C}^{i_n}$ and a basis $B=\{|k_1\dots k_n\rangle\}$ of cardinality $N=i_1\dots i_n$, we define an {\bf experimental arrangement} denoted $\EA_{\Psi,B}^{N,i_1\dots i_n}$, as a specific choice of screens with detectors together with the potentia of each power, that is,
\[
\EA_{\Psi,B}^{N,i_1\dots i_n}= \sum_{k_1,k_1'=1}^{i_1}\dots \sum_{k_n,k_n'=1}^{i_n} 
\alpha_{k_1,\dots,k_n}^{k_1',\dots,k_n'}|k_1\dots k_n\rangle\langle k_1'\dots k_n'|.
\]
The real number $\alpha_{k_1,\dots,k_n}^{k_1,\dots,k_n}$ that accompanies the power $|k_1\dots k_n\rangle \langle k_1\dots k_n|$ is its {\bf potentia} (or intensity) and the basis $B$ determines the powers defined by the specific choice of screens and detectors. The number $N$ which is the cardinal of $B$ is called the {\bf degree of complexity} (or simply degree) of the experimental arrangement. 
Finally, we use {\bf quantum laboratory} (or quantum lab or Q-Lab) as a synonym of ISA. Let us stress that, as we will see, we should not think of a QLab as a classical ``spatial box'' inhabited by quantum powers --as we would tend to do when thinking classically--, we need to think differently.

Given that the extension to a tensorial formulation is possible, and that there are different formal features that need to be considered explicitly, we propose a notation different to the orthodox one: $\EA_{\Psi, B}$ is the experimental arrangement given the ISA, $\Psi$, in the basis $B$ --where the cardinality and the factorization is an information already contained in $B$. This can be extended to $\EA_{\Psi, B}^N$, where  the complexity $N$ makes explicit the number of powers considered, and even more specifically to $\EA_{\Psi, B}^{N, i_1\dots i_n}$ where the numbers $i_1\dots i_n$ make reference to the $i_1$ number of detectors in the first screen, $i_2$ number of detectors in the second screen, $i_3$ number of detectors in the third screen, and so on. In this way it is possible to identify different factorizations $\EA_{\Psi}^{N, i_1\dots i_n}$, $\EA_{\Psi}^{N, i_1'\dots i_n'}$, $\EA_{ \Psi}^{N, i_1''\dots i_n''}$ of the same complexity $N$ without making explicit the basis through the equivalence relation $\EA_{\Psi}^N$ (figure 6).\footnote{Let us remark that Dirac notation tends to confuse the $N\times N$-density matrix $\rho$ in one basis with the equivalence relation $\EA_{\Psi}^N$.}

In the light of these new physical concepts which can be directly related, not only to the mathematical formalism but also, in  operational terms, to experience, we recall the basis and factorization invariance theorems from \cite{deRondeMassri23} as well as their conceptual reading in \cite{deRondeFMMassri23a}:
\begin{theorem}{\sc (Basis Invariance Theorem)}
Given a specific QLab $\Psi$, all experimental arrangements of the same complexity, are equivalent independently of the basis. 
\end{theorem}

\begin{center}
\includegraphics[scale=.35]{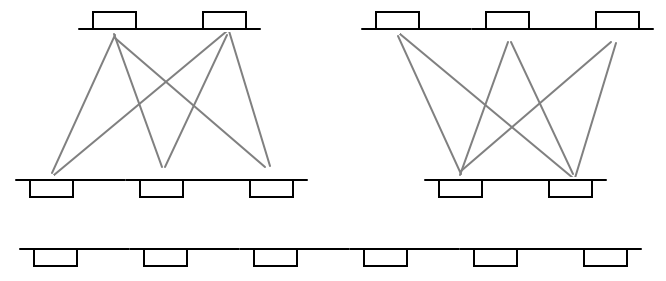}

\small{ \captionof{figure}{Different abstract equivalent representations\\ of the same $\EA_{\Psi}^6$ as $\EA_{\Psi}^{6, 2.3} \equiv \EA_{\Psi}^{6, 3.2} \equiv \EA_{\Psi}^{6, 6}$}}
\end{center}

\begin{theorem} {\sc (Factorization Invariance Theorem)}
The experiments performed within a $\EA_{\Psi}^N$ can also be performed with an experimental arrangement of higher complexity N+M, $\EA_{\Psi}^{N+M}$ that can be produced within the same QLab $\Psi$.  
\end{theorem}

\noindent While the {\it Basis Invariance Theorem} implies that the knowledge we obtain from  the intensities in one experimental arrangement is equivalent to the knowledge obtainable in any other experimental arrangement of the same complexity, the {\it Factorization Invariance Theorem} tells us that constructing an experimental arrangement $\EA_{\Psi}^{N}$ given the knowledge of a more complex experimental arrangement, $\EA_{\Psi}^{N+M}$, will not increase our knowledge in any way. In other words, the knowledge of the experimental arrangement of size $N+M$ will also give us the knowledge of the experimental arrangement of size $N$. We might also remark that {\it the same} power can be part of different experimental arrangements depending on the choice of the basis and factorization. For example, the same power $|i\rangle\langle i|$ can be obtained from the basis 
$B=\{|i\rangle, |j\rangle, |k\rangle\}$ which also determines the powers $|j\rangle\langle j|$ and $|k\rangle\langle k|$, or the basis 
$B'=\{|i\rangle, |r\rangle, |s\rangle\}$ which determines --instead-- the powers $|r\rangle\langle r|$ and $|s\rangle\langle s|$. In general, it is possible to change the basis even though retaining one of the powers by leaving unchanged one of the detectors in each screen (figure 7).
\begin{center}
\includegraphics[scale=.4]{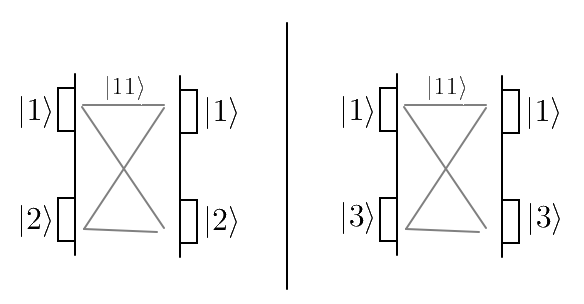}
\small{ \captionof{figure}{Two experimental arrangements with the same power $|11\rangle$.}}
\end{center}

To sum up, given an ISA, the knowledge of a specific EA of degree $N$ implies the knowledge of all EAs in that specific degree of complexity or less. This means that the knowledge of a set of powers (and their intensities) of dimension $N$ will imply the knowledge of the totality of powers (and intensities) within that same dimension or less. A specific experimental arrangement of degree $N$ contains the knowledge of the totality of all possible experimental arrangements of that same complexity. For example, given a Qlab $\Psi$, let us assume that we have two screens with two detectors each. Then, assume that for the basis $B = \{|11\rangle,|12\rangle,|21\rangle,|22\rangle\}$
we have the following experimental arrangement,
\[
\EA_{\Psi,B}^{4,2.2} = \frac{1}{2}|11\rangle\langle 11|+\frac{1}{2}|22\rangle\langle 22|
\]
If we change the detectors in the first screen as $|\uparrow \rangle =( |1\rangle+|2\rangle)/\sqrt{2}$ and $|\downarrow \rangle = (|1\rangle-|2\rangle)/\sqrt{2}$. Then, we have a new basis $B'= \{|\uparrow 1\rangle,|\uparrow 2\rangle,|\downarrow 1\rangle,|\downarrow 2\rangle\}$ and
the experimental arrangement that can be computed by means of linear algebra becomes the following,
\[
\EA_{\Psi,B'}^{4,2.2} = 
\frac{1}{4}|\uparrow1\rangle\langle \uparrow1|+
\frac{1}{4}|\downarrow1\rangle\langle \downarrow1|+
\frac{1}{4}|\uparrow2\rangle\langle \uparrow2|+
\frac{1}{4}|\downarrow2\rangle\langle \downarrow2|+
\frac{1}{2} |\uparrow1\rangle\langle \downarrow1|
- \frac{1}{2} |\uparrow2\rangle\langle \downarrow2|.
\]
In the first case, two of the four powers have intensity $\frac{1}{2}$ and in the second case, the four powers have the same intensity $\frac{1}{4}$. This example emphasizes the fact that the knowledge of one $\EA_{\Psi,B}^N$ implies the knowledge of a different $\EA_{\Psi,B'}^N$. 

Let us also remark it is important not to confuse, as shown in figure 8, one experimental arrangement composed of two screens with two experimental arrangements, each one of them with only one screen:
\begin{center}
\includegraphics[scale=.4]{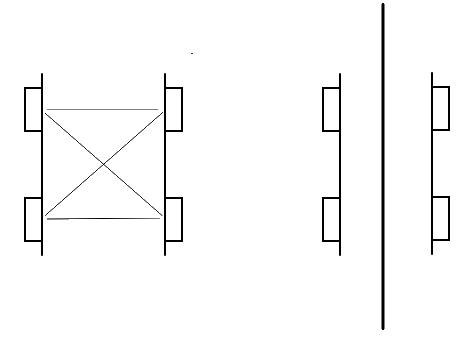}
\captionof{figure}{Difference between $\EA_{\Psi}^{4, 2.2}$ and two $\EA_{\Psi}^{2,2}$.}
\end{center}
This is the difference between one $\EA_{\Psi}^{4, 2.2}$ with 4 powers, 2 screens with two detectors each, and two $\EA_{\Psi}^{2,2}$ with 2 powers each, where each screen has 2 detectors.

The same $\EA_{\Psi}^N$ is thus capable to account consistently for different experimental setups, $\EA_{\Psi, B}^N$, $\EA_{\Psi, B'}^N$, $\EA_{\Psi, B''}^N$, and so on. Notice that the $\EA_{\Psi}^N$ is an abstract expression which has no reference to a particular factorization and basis, and thus cannot be regarded as a meaningful operational physical concept, for as Einstein \cite[p. 26]{Einstein20} stressed: ``The concept does not exist for the physicist until he has the possibility of discovering whether or not it is fulfilled in an actual case.''  In this respect, it should be also stressed that any experimental arrangement presupposes the existence of a particular ISA, say $\Psi$, and a particular basis, say $B$. One cannot refer to an experimental arrangement without a direct reference to a presupposed intensive state of affairs. Thus, the only expression that has operational content and can be considered as a meaningful physical concept is the specific experimental arrangement $\EA_{\Psi, B}^{N, i_1\dots i_n}$. 

\medskip 

So, what is entanglement? According to our analysis it can be understood in the following terms: when, given a QLab (or ISA), an experimental arrangement has 2 or more screens, each power within that EA can be observed as a multiscreen (non-local) effect. This unitary multiscreen effect of powers is what we call quantum entanglement, and it is an inherent feature of every single power. A power --contrary to particles-- can express itself in different screens simultaneously. And alike waves, this phenomena does not involve any ``spooky action at a distance''. However, the reason for this is different. In QM, contrary to wave mechanics, there is no `spatial distance' that can be represented by the mathematical formalism, there is no meaningful way to define the spatial distance between two detectors or screens. Detectors in different screens are represented by vectors in different mathematical spaces, thus, it is impossible to compute any distance between them. The ``distance'' that can be actually computed, via Pitagora's theorem, is that between two powers (or between detectors in the same screen) and it always gives the same result, $\sqrt2$ --which shows it has no physical reference. Hence, the Euclidean distance that could be measured within a lab between two detectors with a ruler simply cannot be represented within the mathematical formalism of the theory. Notice, for example, that the Euclidean distance between two detectors of an experimental set-up constructed within a lab has no effect whatsoever on the quantum phenomena in question. In fact, placing the two detectors in two screens at 1 meter distance will produce exactly the same quantum phenomena as placing them at 1 light-year distance. However, we should also remark that the fact that QM is unable to represent a spatial reality does not mean it cannot make reference to it. To take an analogous example, the electromagnetic theory is able to explain the magnetic effect of two rocks attracting each other even though it is clearly incapable of representing rocks. 

The use of words like ``detector", ``screen", ``lab", etc., can lead to mistaken expectations. Contrary to the Bohrian doctrine of classical concepts, these notions should not be considered as relying on a presupposed classical representation. Even though we might use the same words, in the context of QM these operational physical notions need to be understood internally, with respect to the theory itself, and in this respect, there is no necessary link that must be fulfilled with respect to the representation provided by classical mechanics. As remarked by Heisenberg in an interview by Thomas Kuhn when addressing the creation of QM and the need of new concepts: 
\begin{quotation}
\noindent {\small ``The decisive step is always a rather discontinuous step. You can never hope to go by small steps nearer and nearer to the real theory; at one point you are bound to jump, you must really leave the old concepts and try something new... in any case you can't keep the old concepts.''   \cite[p. 98]{Bokulich06}}
\end{quotation} 
Of course, this does not mean you cannot keep the same words. For example, the words `space' and `time' refer to specific concepts in classical mechanics which differ significantly from the concepts, related to exactly the same words, discussed in relativity theory. This is a consequence of the fact that concepts acquire their meaning in each case from the place they occupy in the specific conceptual system of the particular theory they are part of, this is, their meaning is determined through the relation with the other concepts inside the system. As Heisenberg explained: 
\begin{quotation}
\noindent {\small ``New phenomena that had been observed could only be understood by new concepts which were adapted to the new phenomena. [...] These new concepts again could be connected in a closed system. [...] This problem arose at once when the theory of special relativity had been discovered. The concepts of space and time belonged to both Newtonian mechanics and to the theory of relativity. But space and time in Newtonian mechanics were independent; in the theory of relativity they were connected.'' \cite[pp. 97-98]{Heis58}}
\end{quotation} 

A QLab cannot be represented in terms of a classical Euclidean space where distances can be defined, it is not a ``spatial box" in which classical detectors or screens are situated in different regions of that space. A QLab has nothing to do with classicality. Distances are not something that can be taken into consideration when thinking about a QLab (or when thinking in quantum mechanical terms in general), since such concept has no place within the theory. We can think about the multiplicity of screens and detectors, but not about their spatial distance. In this way, when we say that the same power manifests itself on multiple screens simultaneously, we are not saying that the power travels through space or describes a trajectory --as implied when discussing about quantum particles. We must always avoid the dogmatic projection of classical notions within the theory of quanta, instead, we should produce the understanding of quantum phenomena from the theory itself for, as Einstein explained, ``it is only the theory which decides what can be observed".

The quantum entanglement between detectors is not a ``spooky action at a distance" just in the same way there is nothing ``spooky'' about the magnetic attraction between two rocks --which of course seemed ``spooky'' before the theory of electromagnetism was built. But in the case of powers, we are not even talking about separated entities within a space that is understood as a stage or setting independent of the things it contains. First, because there is not a space independent of the elements considered, and, second, because powers are non-separable. Let us start by the non-spatiality (or non-independence of the medium). The demonstration that powers are not entities within an independent space is straightforward, given the mathematical representation is provided in terms of a {\it configuration space}. Of course, the fact that the theory of quanta implies the existence of a variable {\it multidimensional medium} that depends  on the {\it degrees of freedom} considered marks a radical departure with respect to the most fundamental modern presupposition regarding physical reality, namely, the idea that physical reality must be necessarily restricted to the representation of entities {\it within} a universal, permanent, continuous stage. Perhaps, instead of presupposing a continuous space, we should understand the \emph{medium} entailed by QM in more of a Leibnizian manner,\footnote{As remarked by Arthur: ``Leibniz is celebrated as the most powerful and influential protagonist of the relational theory of space, according to which space consists solely in the relations among bodies, and is not (as Newton claimed) an entity existing in its own right.'' \cite{Arthur13}} where the medium is not an independent universal context in which separated things are placed, but instead, it is produced in a purely relational manner by the powers themselves. In a sense, there is no real separability between the powers and the medium (something that is expressed by the equivalence between the QLab and the ISA). Thus, it is not that powers are “inside” the lab, as one might think of systems within a particular region of space.  Powers are not separated objects that would travel across an independent space. Powers \emph{produce} their medium. The medium itself is nothing but an expression of the powers in any given situation. The ``space" only exists in terms of the interrelation between powers. And without powers there is no ``space". Powers of action produce their own medium --which cannot be understood as --something’ distinct, with a separable existence. 

This leads us to the other novel aspect of QM that we where considering: the non-separability of powers, that we rephrase positively in terms of their intrinsic relationality. Powers are \emph{intrinsically relational}. Powers are not separated independent entities. This means that powers are always multiple, they always come in a relational multiplicity, an ISA is always a multiplicity of related powers. A power alone, by itself, isolated, cannot truly exist. One cannot separate and isolate a single power from the relational whole it is involved in. Powers always exist in relation to other powers, conforming in each case an ISA. Powers constitute the ISA in a relational manner which clearly departs from the way in which our classical reasoning might tend to represent a situation. Powers are not --as in the classical case, when considering particles within a continuous space-- separated elements that populate the lab. Even though they are objective (in the Kantian sense), they are not ``objects'' with an independent reality. Powers of action constitute the QLab in a relational manner.  All powers of action are interconnected and constitute the ISA in a manner which can not be described in terms of a classical set of distinct elements.

This idea of a non-spatial phenomena (i.e., entanglement) and a non-separable representation (i.e., relationalism) was, of course, exactly what Einstein --regardless of his work in relativity theory-- was not ready to accept.\footnote{This might be linked to the fundamental role played by space and time within Kantian modern philosophy as {\it forms of intuition}, different to {\it categorical concepts}.} For him, a necessary precondition for any physical representation was the separation in space as defining the independent existence of individual elements. As he would write in a letter to Max Born dated 5 April, 1948:
\begin{quotation}
\noindent {\small ``If one asks what, irrespective of quantum mechanics, is characteristic of the world of ideas of physics, one is first stuck by the following: the concepts of physics relate to a real outside world, that is, ideas are established relating to things such as bodies, fields, etc., which claim a `real existence' that is independent of the perceiving subject --ideas which, on the other hand, have been brought into as secure a relationship as possible with the sense-data. {\it It is further characteristic of these physical objects that they are thought of as arranged in a space-time continuum. An essential aspect of this arrangement of things in physics is that they lay claim, at a certain time, to an existence independent of one another, provided these objects `are situated in different parts of space'.} Unless one makes this kind of assumption about the independence of the existence (the `being-thus') of objects which are far apart from one another in space --which stems in the first place in everyday thinking-- physical thinking in the familiar sense would not be possible. It is also hard to see any way of formulating and testing the laws of physics unless one makes a clear distinction of this kind.'' \cite[p. 170]{Born71} (emphasis added)}
\end{quotation}
However, as Einstein himself recognized: 
\begin{quotation}
\noindent {\small ``There seems to me no doubt that those physicists who regard the descriptive methods of quantum mechanics as definite in principle would react to this line of thought in the following way: they would drop the requirement for the independent existence of the physical reality present in different parts of space; they would be justified in pointing out that \emph{the quantum theory nowhere makes explicit use of this requirement}.'' \cite[p. 172]{Born71}} (emphasis added)
\end{quotation}
And this is indeed the case. We might say that powers of action are intrinsically non-spatial --at least not in the classical sense-- and non-separable.

\section{Everything is Entangled in Quantum Mechanics}

We have shown how entanglement, as the unitary multiscreen effect of a single power, is an irreducible aspect of the operational content of the theory of quanta. The theory talks about powers of action each one them producing a multiscreen (non-local) effect that can be observed in the lab. Consequently, there is nothing non-entangled in QM. There is no meaningful distinction between something that is entangled and something that is not entangled within the theory of quanta. The attempt of quantifying or measuring the level of entanglement becomes meaningless. This idea is of course a direct consequence of imposing the dogmatic metaphysical presupposition according to which there must exist a ``correspondence'' or ``limit'' between the micro and the macro, between the quantum and the classical. However, as it is well known, after more than a century of Bohr's introduction of the {\it correspondence principle} in 1913 \cite{BokulichBokulich20}, there has been no success within physics in providing a representation of such supposedly existent bridge between the quantum and the classical. In fact, the model of decoherence which became popular during the 1980s as a solution to the measurement problem and the quantum to classical limit was soon exposed --due to the many formal and conceptual inconsistencies found within the specialized philosophical literature-- as a complete failure.\footnote{unfortunately, turning things upside-down, this failure was then presented as a new type of ``solution for all practical purposes'' (in short, a ``FAPP solution''). A new form of the famous ``It works. So shut up and calculate!''} So what can we say about the quantification of entanglement? According to our analysis, when focusing in the formalism of QM itself --quite regardless of the classical representation--, the measure that does provide an interesting set of consequences is the {\it measure of complexity} of the EAs. This specific measure describes the original {\it relations of equivalence} between the EAs of the same degree of complexity (an EA of a certain degree of complexity allows you to derive all other EAs of that degree with all their powers) as well as a chain of {\it complexity inclusions} between EAs of different degrees which, in turn, allow us to quantify relations between powers in the following manner: 
\[
\EA_{\Psi}^1 \subsetneq \EA_{\Psi}^2 \subsetneq \dots \subsetneq 
\EA_{\Psi}^N\subsetneq \dots \subsetneq \EA_{\Psi}^\infty
\]
where $\EA_{\Psi}^1$ is the trivial experimental arrangement with only one power and $\EA_{\Psi}^\infty$ represents an experimental arrangement with an infinite number of powers. Also, according to the previous theorems, while $\EA_{\Psi}^1$ is the experimental arrangement with {\it minimal complexity}, namely, the experiment which has only information about powers of degree of complexity 1; $\EA_{\Psi}^\infty$ is the experimental arrangement with {\it maximal complexity}, the one with the complete information of all possible experimental arrangements and thus of all possible powers and potentia that might be considered within that ISA.

The {\it degree of complexity} (i.e., the number of powers of action) of an experimental set up, also given by the dimension of the {\it configuration space} in which $\Psi$ is represented, also provides a quantification of the \emph{knowledge} we possess of an Intensive State of Affairs. Let us remark that in analogous terms to the classical case, possessing the knowledge of an ISA  in QM (which is analogous to possessing the knowledge of an ASA in the classical case) implies having a complete account of that particular state of affairs, and thus \emph{maximal knowledge}. This result stands in extreme contraposition to the orthodox account where it is the (operational) definition of {\it pure state} which provides {\it maximal knowledge} of a given situation --in this case with respect to single measurement outcomes. It is essential to stress that while in the first case the complete knowledge refers to the ISA, in the latter orthodox case it refers to single measurement outcomes (or actual observations). This empiricist reference, translated to our scheme, is described by an $\EA_{\Psi}^1$, namely, a single screen with only one detector, which in our case is the {\it minimum knowledge} we can get of the state of affairs from an experimental set up.

\section*{Acknowledgements} 

This work was partially supported by the following grants:  ANID-FONDECYT, Project number: 3240436.

\end{document}